\documentstyle[12pt]{article}
\begin{document}



\begin{center}{\Large\bf Vanishing of Gravitational Particle Production}
\end{center}
\begin{center}
{\Large\bf
 in the}\end{center}
\begin{center}{\Large\bf
Formation of Cosmic Strings}
\end{center}

\bigskip

\begin{center}
{\large Iver Brevik}\footnote{
e-mail:iver.h.brevik@mtf.ntnu.no}\\
\smallskip
{\it Division of Mechanics, Norwegian University of Science and Technology,\\
N-7034 Trondheim, \\
Norway}
\end{center}

\smallskip

\begin{center}
{\large Bj\o rn Jensen}\footnote{e-mail:bjensen@gluon.uio.no}\footnote{
On leave of absence from Institute of Physics, University of Oslo, Norway.}\\
\smallskip
{\it NORDITA,\\
Blegdamsvej 17, DK-2100 Copenhagen \O ,\\
Denmark}
\end{center}


\bibliographystyle{unsrt}


\bigskip

\begin{abstract}
We consider the gravitationally induced
particle production from the quantum vacuum which is defined
by a free, massless and minimally coupled scalar field during the formation
of a gauge cosmic string. 
Previous discussions of this topic estimate the power output per
unit length along the string to be of the order of $10^{68}$ ergs/sec/cm in the 
s-channel.
We find that this production may be completely suppressed. A similar
result is also expected to hold for the number of produced photons.

\end{abstract}

\begin{center}
{\bf PACS} number(s): 11.27.+d, 98.80.Cq
\end{center}

\newpage

\section{Introduction}

When a long and straight gauge cosmic string is formed, it is expected that the 
resulting
spacetime geometry outside the string core
to a very high degree of approximation can be described by a flat space
with a conical deficit angle. Previous studies indicate that the formation of such
strings is accompanied by an intense burst of particles and radiation which are
released from the quantum vacuum.
The power output per unit length along a string during the formation process may be
very large as to be
of the order of $10^{68}$ ergs/sec/cm for a free, massless and minimally coupled
scalar field \cite{Parker}. The power output is strongly suppressed
when one considers higher angular momentum quantum numbers. 
A similar estimate for the production of photons was deduced in \cite{Brevik}.
The purpose
of this communication is to show that the energy produced during the formation of 
a gauge cosmic string may be significantly lower than what previous estimates have 
indicated.
It is found in particular that there is no energy production in 
the zero-angular momentum sector when one
considers the free, massless and minimally coupled scalar-field sector to leading
order in the string tension. We also expect a similar conclusion to hold for the
number of produced photons.

\section{A String Geometry}

Let us first briefly review some properties of the geometry which is induced 
by a gauge cosmic string.
It is in general believed
that the spacetime geometry outside a long and straight gauge cosmic string can
be approximated by the spacetime found outside a corresponding fundamental string,
since the typical radius of a cosmic string is typically of the order of $10^{-30}$ cm..
Hence, the cosmic string action can be approximated by the fundamental string action
\begin{equation}
S=-\frac{T}{2}\int d^2\sigma\sqrt{-h}\partial_\mu X^A\partial_\nu X^Bh^{\mu\nu}g_{AB}\, 
,
\end{equation}
where Greek letters denote world-sheet indices, while capital Latin ones denote 
target-space
coordinates. $h$ denotes the world-sheet geometry, $g$ the target space geometry, and
$T\, (>0)$ is the string tension. 
It was found in \cite{Jensen1,Jensen2} that such a string 
which is static, straight and infinitely long will, in the {\it Weyl gauge} and 
using cylinder coordinates, 
give rise to the geometry
\begin{equation}
ds^2=a_0^{-1}(-dt^2+dr^2+dz^2)+a_0r^2d\phi^2
\end{equation}
where $a_0=1-4G\mu_0$, and $\mu_0(=T)$ is the proper energy density in the string.
In the following
we will use both $\mu_0$ and $T$ to denote the proper energy density in the string.
It is at the outset put no restrictions on the value of $\mu_0$ so that
$\mu_0$ parametrises a {\em single} and infinitely large family
of geometries. 
The string source is assumed
to be positioned at $r=0$, such that the region $r>0$ is a flat vacuum region.
The coordinates are all assumed to be independent of each other, and of $\mu_0$.
The world-sheet geometry is related to the target space geometry by
\begin{equation}
h_{\mu\nu}=\partial_\mu X^A\partial_\nu X^B g_{AB}\, .
\end{equation}
By adjusting the $t$ and $z$ coordinates to coincide with the timelike and spacelike
coordinates in the string world-sheet we have that
\begin{equation}
h_{\mu\nu}=a_0^{-1}\eta_{\mu\nu}\, .
\end{equation}
Clearly, when $\mu_0>(4G)^{-1}$ we see that the world-sheet geometry changes sign,
while the orientation of the induced geometry in the string world-sheet is unchanged.
It is correspondingly straightforward  
to see that the target space geometry changes signature when
$\mu_0$ becomes greater that $(4G)^{-1}$.
The geometry in eq.(2) can be brought in the form
\begin{equation}
ds^2=-dT^2+dR^2+dZ^2+a_0^2R^2d\phi^2
\end{equation}
with the use of the rescalings $T=a_0^{-1/2}t$, $R=a_0^{-1/2}r$ and $Z=a_0^{-1/2}z$
{\em provided that} $\mu_0<(4G)^{-1}$.
This last expression for the string geometry can be derived directly from
the Levi-Civita form of the line-element \cite{Hiscock,Jensen2}.

The action in eq.(1) is not invariant under
target space diffeomorphisms. Using eq.(5) we find that the string action takes the form
\begin{equation}
S=-\frac{T}{2}\int d^2\sigma\partial_\mu X^A\partial_\nu X^B\eta^{\mu\nu}\eta_{AB}\, ,
\end{equation}
where $\eta$ denotes the corresponding Minkowski geometry, while using the geometry in 
eq.(2)
we find that
\begin{equation}
S=-\frac{T}{2a_0}\int d^2\sigma\partial_\mu X^A\partial_\nu X^B\eta^{\mu\nu}\eta_{AB}\, 
.
\end{equation}
Hence, in the geometry eq.(2) we can perceive the string to carry a renormalised tension 
$T_{\mbox{\small ren}}$which
is given by \footnote{One can also come to this conclusion without the explicit use of 
the
string action by a very simple and straightforward calculation of the energy density
in the string
in the coordinate system in eq.(2), when $T$ is defined as the proper energy density
in the string.}
\footnote{Note that this string tension renormalisation is of a purely classical
and non-perturbative nature, and does therefore represent an additional string
tension renormalisation mechanism in addition to the perturbative ones of quantum
mechanical origin (see \cite{Damour}, and the references therein).}
\begin{equation}
T_{\mbox{\small ren}}\equiv\frac{T}{1-4GT}\Rightarrow 
T=\frac{T_{\mbox{\small ren}}}{1+4GT_{\mbox{\small ren}}}\, .
\end{equation}
Since $\mu_0$ (or $T$) denotes the canonical energy measure of the string relative
to the geometry in eq.(5), it is more correct to formulate the geometry in eq.(2)
in terms of the energy measure of the string relative to the coordinate system which is
used there, i.e. in terms of $T_{\mbox{\small ren}}$. The string geometry 
then takes the form
\begin{equation}
ds^2=b_0(-dt^2+dr^2+dz^2)+b_0^{-1}r^2d\phi^2\, ,
\end{equation}
where $b_0\equiv 1+4GT_{\mbox{\small ren}}=(1-4G\mu_0)^{-1}\geq 1$.

 In \cite{Parker}, and in a large number of consecutive
studies (see \cite{Brevik,Pullin,Mendel}, e.g.), 
the induced string geometry was taken in the form
\begin{equation}
ds^2=-dt^2+dr^2+dz^2+a_0^2r^2d\phi^2\, .
\end{equation}
We have used the {\em same} coordinates in eq.(10)
as in the other equations above. \footnote{The reason for identifying the coordinates in
eq.(2) and eq.(10) is to be able to relate the number of particles
released in string formation when calculated in the geometry in eq.(2) and eq.(10),
since quantum field theory is {\em not} generally covariant.}
We can do this
since it is assumed (implicitly) in all the references above that the coordinates
used in eq.(10) 
\begin{flushleft}{\bf (1)} cover the complete spacetime manifold ($r>0$)\end{flushleft}
\begin{flushleft}{\bf (2)} do not depend on $\mu_0$. \end{flushleft}
The singularity structure carried by the manifold described by eq.(10)
coincide with the singularity structure in the manifold described in eq.(2) at $r=0$
(a curvature singularity), but the
singularity structures differ significantly at the point $\mu_0=(4G)^{-1}$ in
the $\mu_0$-parameter space where the geometry in eq.(2) changes
signature and $T_{\mbox{\small ren}}\rightarrow\infty$. 
Hence, since eq.(2) is derived with no particular restriction on the value
of $\mu_0$
eq.(2) and eq.(10) describe two {\em different} (families of)
manifolds. However, at each {\em fixed} point $\mu_0$ in the $\mu_0$-parameter space,
and such that $0\leq\mu_0<(4G)^{-1}$ (which is the physically interesting regime),
these spaces can be brought into manifestly corresponding forms via the rescalings
following eq.(5). These spaces can also be so related when $\mu_0>(4G)^{-1}$. However, 
in this
case we must in addition to a set of rescalings also change the signature of either the 
geometry
in eq.(10), or the geometry in eq.(2). Note that $a_0$ is the same in both eq.(2)
and eq.(10) since it is a function of the {\em proper} energy density in the source.
One way to give a presice description of the relation between the geometry in eq.(2)
and the geometry
in eq.(10) is to observe \cite{Jensen1,Jensen2} that eq.(2) can be written in the form
\begin{equation}
ds^2=a_0^{-1}(-dt^2+dr^2+dz^2+a_0^2r^2d\phi^2)\, .
\end{equation}
Hence, {\em under the assumptions in {\bf (1)} and {\bf (2)} above},
it follows that the
geometry which is used in a majority of previous works
on gauge cosmic string theory is conformally related to the geometry in eq.(2), i.e. one
can go from eq.(2) and to the form in eq.(10) by multiplication with an overall
constant scale factor.

\section{Gravitational Particle Creation}

Even though a cosmic string may have the impressing proper mass per unit length
$\mu_0\sim 10^{22}$ g/cm $\sim 10^{43}$ erg/cm, the deviation from the Minkowski space
which the string induces in $dt=dz=0$ hyperplanes is only of the (less impressing)
order $G\mu_0\sim 10^{-6}$. However,
it has been argued in earlier studies that
even though a static string barely distorts spacetime, the energy production $W$ during 
the
formation of the string may be huge, and at least 
 of the order of $W\sim 10^{20}$ erg/cm during a formation
time which was taken to be
of the order $\Delta t\sim 10^{-35}$ sec.. 
In these studies a variety of different scalar fields as well
as Maxwell fields were studied. Hence, 
physical particles released from the quantum vacuum due to the
creation of a string
may represent a potentially significant source of entropy in the early universe 
\cite{Parker}.
We will now turn to a reassessment of 
these production estimates.

In the computation of the energy released in the creation of a string we will
use the instantaneous approximation, which was pioneered in this context in 
\cite{Parker}.
In this approach one assumes that the creation of the string happens instantaneously.
We will follow \cite{Parker} in that we will
assume that the initial spacetime is Minkowski space,
but we will describe the final spacetime by eq.(9) (or equivalently eq.(11))
and not by eq.(10). The actual time-dependent
problem can thus
be captured by writing the relevant spacetime geometry as in eq.(9), but
with $T_{\mbox{\small ren}}$ replaced by a time-dependent 
function. In our case this time-dependent function
should be the Heaviside step-function $\Theta$,
i.e.
\begin{equation}
T_{\mbox{\small ren}}\rightarrow T_{\mbox{\small ren}}(t)=T_{\mbox{\small ren}}\Theta 
(t-t_0)
\Rightarrow b_0\rightarrow b(t)\, .
\end{equation}
The moment of creation of the string is at $t=t_0$. 
In explicit calculations we will set $t_0=0$ without loss of any generality.

We will direct our attention
to a massless and minimally coupled scalar field $\Phi (x)$ configuration with the 
density
\begin{equation}
{\cal L}=\sqrt{-g}\partial_A\Phi (x)\partial^A\Phi (x)\, .
\end{equation}
Since we always have that 
$\sqrt{-g}g^{AA}=1\, ;\, A\neq\phi$, it follows that the corresponding field equation 
reduces
to the form
\begin{equation}
(\Box_3+\frac{b^2(t)}{r}\partial^2_\phi)\Phi (x)=0\, ,
\end{equation}
where $\Box_3$ is the d'Alembertian in a 2+1-dimensional Minkowski geometry
for {\it all times} $t$.
The reader should be aware that the form of eq.(14) does not
depend on the particular substitution in eq.(12), but is valid
for {\em any} general function $b(t)$. 
From this
equation of motion it follows in particular
that the continuity condition across $t=t_0$ in the time direction simply reduces to
\begin{equation}
(\partial_t\Phi (x))|_{t_0^{-}}=(\partial_t \Phi (x))|_{t_0^{+}}\, .
\end{equation}
This fact will be of crucial importance, and will represent one main reason, for why
our findings differ from those stemming from similar previous excursions into this 
subject.

In the following we will confine the quantum field to the interior of a straight
cylinder centered at the origin $r=0$, with the constant coordinate radius $r=R$.
The top and buttom of the cylinder are assumed to be at
the fixed coordinate positions $z=0$ and $z=L$, respectively.
The cylinder thus defines a {\em co-moving} volume, since the proper volume
changes in the transition from Minkowski spacetime and to the situation when a 
string is present. Note that the proper area of a cross section of the
cylinder defined by $dt=dz=0$ always equals $\pi R^2$. The change in the proper
volume of the cylinder is thus solely induced by a ``stretching'' of the cylinder in
the $z$-direction.

In the region $t<t_0$ we decompose the scalar field operator according to
\begin{equation}
\Phi (\vec{x},t)=\sum_{j}( a_jf_j(\vec{x},t)+a^\dagger_jf^*_j(\vec{x},t))\, ,
\end{equation}
where $j=(n,m,s); n,m=0,\pm 1,\pm 2,..., s=1,2,3,...$, and the annihilation and the 
creation
operators $a_j$ and $a^\dagger _j$ 
satisfy the usual canonical commutation relations. 
The mode functions
$f_j$, which constitute a complete set with respect to the canonical symplectic form,
are given by
\begin{equation}
f_{n,m,s}(\vec{x},t)=N_1e^{-i\omega_st}e^{ikz}e^{im\phi}J_{|m|}((\omega_s^2-k^2)^{1/2}r)
\, ,
\end{equation}
where the normalisation factor $N_1$ is given by
\begin{eqnarray}
&&N_1=(2\omega_s V_1)^{-1/2}((\partial_rJ_{|m|} ((\omega_s^2-k^2)^{1/2}r)|_{r=R})^{-1}\, 
,\\
&&V_1\equiv\pi LR^2\,\, ,\,\, k=\frac{2\pi n}{L}\, .
\end{eqnarray}
The $\omega_s$'s are determined from the equation $J_{|m|}((\omega_s^2-k^2)^{1/2}R)=0$.
$s$ does therefore denote a radial quantum number.
The canonical vacuum state is defined as $a_j|0\rangle_{\mbox{\small in}}=0$.
When $t>t_0$ we similarly expand the quantum field as
\begin{equation}
\Phi (\vec{x},t)=\sum_{j}(b_jg_j(\vec{x},t)+b^\dagger _jg^*_j(\vec{x},t))\, .
\end{equation}
The mode-functions $g_j$ are given by
\begin{equation}
g_j(\vec{x},t)=N_2e^{-iW_st}e^{ikz}e^{im\phi}J_{\nu_m}((W_s^2-k^2)^{1/2}r)
\end{equation}
with $\nu_m\equiv b_0 |m|$. 
The $W_s$'s are determined from $J_{\nu_m}((W_s^2-k^2)^{1/2}R)=0$. The canonical
symplectic form is defined by
\begin{equation}
(\psi_n,\psi_m)=i\int_{\Sigma}d^3x\sqrt{g_3}N^A\psi_n{\mathop{\partial}
\limits^{\leftrightarrow}} _{_A}\psi^*_m\, ,
\end{equation}
where $\vec{N}$ is defined as a future pointing unit vector, $\vec{N}^2=-1$, which is 
everywhere
orthogonal to the spacelike hypersurface $\Sigma$. 
$d^3x\equiv drdzd\phi$ and $g_3$ is the determinant of the
induced 3-geometry in $\Sigma$. 
We choose the normal vector field to be the canonical one, i.e. we choose
\begin{equation}
\vec{N}|_{t_0^-}=\partial_t\,\, ,\,\, \vec{N}|_{t_0^+}=b_0^{-1/2}\partial_t\, .
\end{equation}
With this normalisation we find that
\begin{equation}
(g_j,g_j)=i\int_{\Sigma} d^3x r(g_j
\mathop{\partial_t}\limits^{\leftrightarrow} 
g^*_j)|_{t_0^+}\, .
\end{equation}
It follows that $N_2$ has the same form as $N_1$ except that $\omega\rightarrow W$
and $|m|\rightarrow \nu_m$,
of course.

On the spacelike
hypersurface $\Sigma$ of instantaneous creation of the 
string we will have the following
general relation between $f$-modes and $g$-modes
\begin{equation}
f_j(\vec{x},t_0^-)=\sum_{j'}(\alpha_{j';j}g_{j'}(\vec{x},t_0^+)
+\beta_{j';j}g^*_{j'}(\vec{x},t_0^+))\, .
\end{equation}
The total number of produced particles $|\beta |^2$ 
as they are defined in the $t>t_0$ region
relative to the incoming vacuum $|0\rangle_{\mbox{\small in}}$ is formally given by
\begin{equation}
|\beta |^2\equiv\sum_{j'j}|\beta_{j';j}|^2\equiv
\sum_{j'j}\, _{\mbox{\small in}}\langle 0|N_{j'j}|0\rangle_{\mbox{\small in}}\, ,
\end{equation}
where $N_{j'j}\equiv b^\dagger _{j'}b_j$. From eq.(25) we then easely deduce that
\begin{eqnarray}
\beta_{j';j}&=&i\int_\Sigma d^3x(f_j|_{t_0^-}(\sqrt{g_3}\vec{N}g^*_{j'})|_{t_0^+}-
g_{j'}|_{t_0^+}(\sqrt{g_3}\vec{N}f^*_j)|_{t_0^-})\\
&=&i\int_\Sigma d^3x(f_j|_{t_0^-}(\partial_t g^*_{j'})|_{t_0^+}-
g_{j'}|_{t_0^+}(\partial_tf_j^*)|_{t_0^-}))\, .
\end{eqnarray}
Since we have 
\begin{equation}
f_{n,0,s}|_{t_0^-}=g_{n,0,s}|_{t_0^+}\,\, ,\,\, 
\partial_tf_{n,0,s}|_{t_0^-}=\partial_tg_{n,0,s}|_{t_0^+}\, , 
\end{equation}
we can immediately conclude that
\begin{equation}
\beta_{n,0,s;n,0,s}=0 \, . 
\end{equation}
However, even though the diagonal elements in the scattering matrix $\beta_{s';s}$
vanish, the off-diagonal elements may not. The explicit form for $\beta_{j';j}$ is
\begin{equation}
\frac{\beta_{j';j}}{\beta^{(10)}_{j';j}}=b_0^{1/2}(\frac{2(s'-s)-|m|+|m'|b_0}{(2s'+\frac
{3}{2})
+(|m'|-|m|-2s-\frac{3}{2})b_0})\, ,
\end{equation}
where $\beta^{(10)}_{j';j}$ represents the resulting
production estimate if we had used the
geometry in eq.(10) in order to compute the particle production.
We identify $\beta^{(10)}_{j';j}$ with the corresponding expression in \cite{Parker}.
In the derivation  of 
this expression (which is straightforward and will therefore not be reproduced here) 
we assumed that $R$ is very large in order to utilise
the asymptotic properties of the Bessel functions. We also put $n=n'=0$, since
the inclusion of these quantum numbers does not provide us with any additional insights.
In the physically interesting regime we can effectively set $m=m'=0$ \cite{Parker}, 
so that
\begin{equation}
\frac{\beta_{s';s}}{\beta^{(10)}_{s';s}}=(\frac{2(s'-s)\sqrt{1-4G\mu_0}}
{((2s'+\frac{3}{2})(1-4G\mu_0)-(2s+\frac{3}{2}))})\, .
\end{equation}
Clearly, $\beta_{s;s}=0$, while $\beta_{s';s\neq s'}\neq 0$. However, this
off-diagonal
production is sub-leading. Indeed, from \cite{Parker} one finds that to leading order
in $4G\mu_0$($<<1$)
\begin{equation}
\beta^{(10)}_{s';s}\sim 2G\mu_0\delta_{s',s}\, .
\end{equation}
Hence, the potentially {\em physically significant}
production of scalar particles from the quantum vacuum, due to the creation
of a physically realistic cosmic string, 
is completely suppressed in the zero angular momentum sector.

\section{Conclusion}

In previous studies \cite{Parker,Brevik,Pullin,Mendel} (e.g.)
one made the replacement
$a_0\rightarrow a(t)$ in eq.(10)
in order to approximate the description of cosmic string creation.
Clearly, the resulting geometric
structure is not simply related to the corresponding geometry in
eq.(2). From this point of view it is perhapse not surprising that our
results differ significantly from the results of these previous studies.
The conformal form of the string geometry in eq.(11) were explored and partially
utilised in \cite{Jensen1,Jensen2} 
in order to understand the properties of the usual form of the
string geometry in eq.(10). In \cite{Jensen3} this form 
of the string geometry
was also used in order to extract
the qualitative properties of the amount of particles produced during string creation
compared to the estimate one computes directly from eq.(10). When the conformal
scalar field sector is considered along with the associated conformal vacuum structure
\cite{Birrell}, it was shown in \cite{Jensen3}
that the total number of particles 
produced in the appropriately generalised version of eq.(11) ($|\beta^{(11)}|^2$)
is less than the corresponding amount produced in eq.(10) ($|\beta^{(10)}|^2$),
in the instantaneous approximation. These quantities 
were found to be related by \cite{Jensen3}
\begin{equation}
|\beta^{(11)}|^2=(1-4G\mu_0)|\beta^{(10)}|^2\, .
\end{equation}
Clearly, the conformal vacuum is a very special configuration, and is probably of 
greater
theoretical interest than of physical significance. 
However, the findings in
\cite{Jensen3} are indeed along the lines implied by the findings in this paper,
since $|\beta^{(11)}|^2<|\beta^{(10)}|^2$ to leading order in $4G\mu_0$.
Indeed, the $(1-4G\mu_0)$ factor in eq.(34) is also manifestly present
in the squared version of eq.(32).
However, the differences between eq.(34) and eq.(32) 
do also illustrate that the question of
whether any 
physically significant particle production occur in string creation or not,
is very sensitive to the nature of the coupling of the scalar field to gravity.


\section{Acknowledgements}

BJ thanks NORDITA for a travelling grant, and NORDITA, 
the Norwegian University of Science and Technology
and the Niels Bohr Institute
for hospitality during the time when this work was carried out.

\end{document}